\documentclass[twocolumn]{aastex63}

\newcommand{\pcm}{\,pc cm$^{-3}$}	

\received{XXX}
\revised{XXX}
\accepted{XXX}
\submitjournal{ApJ}

\shorttitle{Vela giant micropulse emission at C band}
\shortauthors{J. L. Chen et al.}

\begin{document}

\title{Giant micropulse emission in the Vela pulsar at C band}

\correspondingauthor{Z. G. Wen}
\email{wenzhigang@xao.ac.cn}

\author{J. L. Chen}
\affiliation{Department of Physics \& Electronic Engineering, \\
Yuncheng University, Yuncheng, Shanxi, 044000, China}

\author{Z. G. Wen}
\affiliation{Xinjiang Astronomical Observatory, Chinese Academy of Sciences, \\
150, Science-1 Street, Urumqi, Xinjiang, 830011, China}
\affiliation{Key laboratory of Radio Astronomy, Chinese Academy of Sciences, \\
Nanjing, 210008, China}

\author{L. F. Hao}
\affiliation{Yunnan Astronomical Observatory, Chinese Academy of Sciences, \\
Kunming, 650011, China}

\author{J. P. Yuan}
\affiliation{Xinjiang Astronomical Observatory, Chinese Academy of Sciences, \\
150, Science-1 Street, Urumqi, Xinjiang, 830011, China}
\affiliation{Key laboratory of Radio Astronomy, Chinese Academy of Sciences, \\
Nanjing, 210008, China}

\author{J. Li}
\affiliation{Xinjiang Astronomical Observatory, Chinese Academy of Sciences, \\
150, Science-1 Street, Urumqi, Xinjiang, 830011, China}
\affiliation{Key laboratory of Radio Astronomy, Chinese Academy of Sciences, \\
Nanjing, 210008, China}
\affiliation{Key laboratory of Microwave Technoloty, Urumqi, Xinjiang, \\
830011, China}

\author{H. G. Wang}
\affiliation{School of Physics and Electronic Engineering, \\
Guangzhou University, 510006 Guangzhou, PR China }
\affiliation{Xinjiang Astronomical Observatory, Chinese Academy of Sciences, \\
150, Science-1 Street, Urumqi, Xinjiang, 830011, China}

\author{W. M. Yan}
\affiliation{Xinjiang Astronomical Observatory, Chinese Academy of Sciences, \\
150, Science-1 Street, Urumqi, Xinjiang, 830011, China}
\affiliation{Key laboratory of Radio Astronomy, Chinese Academy of Sciences, \\
Nanjing, 210008, China}

\author{K. J. Lee}
\affiliation{Kavli Institute for Astronomy and Astrophysics, Peking University, \\
Beijing, 100871, China }

\author{N. Wang}
\affiliation{Xinjiang Astronomical Observatory, Chinese Academy of Sciences, \\
150, Science-1 Street, Urumqi, Xinjiang, 830011, China}
\affiliation{Key laboratory of Radio Astronomy, Chinese Academy of Sciences, \\
Nanjing, 210008, China}

\author{Y. H. Xu}
\affiliation{Yunnan Astronomical Observatory, Chinese Academy of Sciences, \\
Kunming, 650011, China}

\author{Z. X. Li}
\affiliation{Yunnan Astronomical Observatory, Chinese Academy of Sciences, \\
Kunming, 650011, China}

\author{Y. X. Huang}
\affiliation{Yunnan Astronomical Observatory, Chinese Academy of Sciences, \\
Kunming, 650011, China}

\author{R. Yuen} 
\affiliation{Xinjiang Astronomical Observatory, Chinese Academy of Sciences, \\
150, Science-1 Street, Urumqi, Xinjiang, 830011, China}

\author{M. Mijit}
\affiliation{School of Physics, Xinjiang University, \\
Urumqi 830046, China }

\begin{abstract}
We present here the analysis of giant micropulses from the Vela pulsar.
A total of 4187 giant micropulses with peak flux density $>$2.5 Jy were detected 
during almost 4 hours of observations carried out with the Yunnan 40-m radio 
telescope at 6800 MHz.
Nine of the giant micropulses arrived approximately 3 to 4 ms earlier than the peak
of average pulse profile, longer than that at lower frequencies.
The remaining giant micropulses were clustered into three distributions which
correspond to three main emission regions, including four occurring on the
trailing edge of averaged profile.
We find that the peak flux density distribution follows a power law with index
$\alpha \approx -4$.
Furthermore, a certain amount of memory is present from the giant micropulse waiting time
distribution.
Possible emission mechanisms are discussed.
\end{abstract}


\section{Introduction}
\label{sec:intro}
Vela pulsar (PSR J0835$-$4510 or B0833$-$45) is a multi-wavelength emitting, 
young, close, luminous and isolated neutron star associated with the 
Vela supernova remnant in the constellation of Vela.
It is known to emit giant micropulses with high peak flux density and 
narrow pulse width, which are located at the leading edge of the pulse 
profile both at 660 MHz and 1413 MHz \citep{Johnston+etal+2001}.
No genuine giant pulses have been detected since their mean flux densities do
not exceed 10 times the mean flux density of average pulse profile, according 
to the working giant pulse definition \citep{Knight+2006}.
The giant micropulses at 2.3 GHz were observed to have a
power-law distribution of flux density with a slope of $-2.85$
\citep{Kramer+etal+2002}.
Consecutive bright radio pulses with five times the flux of the average pulse were 
detected at 1440 MHz, which suggests that the individual bright pulses may not be 
independent random events \citep{Palfreyman+etal+2011}.

At lower observing frequencies, the pulse scatter broadening smooths over the
microstructure features, and the pulse intensity fluctuates in time and
frequency domains caused by the interstellar scintillation.
These effects can be ruled out at higher frequencies.
Nonetheless, no giant micropulses above 2.3 GHz have been mentioned in the literatue.
Furthermore, whether giant micropulses are limited in the leading edge of the
average pulse profile, and whether they evolve with observing frequencies, are
necessary to be investigated.
No strict naming convention for giant micropulses has been formalized.
In this paper, we present results of giant micropulse emission (peak flux
density $>$2.5 Jy) from the Vela pulsar at higher frequency.
The nomenclature of giant micropulse will be used here for continuty.
In Section~\ref{sec:obs}, we describe our observations and data reduction.
We show the results on giant micropulse emission in Section~\ref{sec:results}.
In Section~\ref{sec:discussion}, we discuss our findings.

\section{Observations and Data Reduction} 
\label{sec:obs}
The Vela pulsar was observed on 17 August, 2019 using the Yunnan 40-m radio 
telescope in a frequency band centered at 6800 MHz.
The orthogonal linear polarizations with a bandwidth of 800 MHz ($\sim$508 MHz
is usable) were injected into a cryogenic receiving system.
Then the output power was recorded with 1024 channels over the passband using 
a ROACH2\footnote{https://casper.berkeley.edu/wiki/ROACH2} based digital filterbank 
system with an effective sampling time of 40.96 $\mu$s.

Observations of individual pulses from the vela pulsar allowed us to perform a variety 
of analysis techniques which we describe below.
Before performing the analysis, data were converted into the filterbank 
format required for the SIGPROC\footnote{http://sigproc.sourceforge.net/} 
analysis package, and only the total intensity (Stokes I) was preserved.
The radio frequency interference (RFI) was rejected by excluding narrow band as
well as bursty broad band RFI by visual inspection.
Incoherent de-dispersion was performed to remove sub-channel dispersive smearing
using a dispersion measure (DM) of 67.97\pcm \citep{Petroff+etal+2013}.
Then, the data were folded to 2182 phase bins across the period with the ephemery 
of the pulsar, using the TEMPO\footnote{http://tempo.sourceforge.net/} package
to obtain a single pulse sequence for further analysis.
In total, almost 4-hour successive observations we observed over 150,000
rotations of the pulsar.

Subsequently, the temporally resolved pulses were converted to flux based on a
nominal system temperature of 40 K and efficiency of 50\% for the C band
receiver, since no flux calibrators were observed.
With these values, the average flux density is measured to be
$7.3\pm1.8$ mJy by integrating intesity of the folded pulse profile over the
entire period.
Generally, the pulsar spectra follow a simple power law 
$S_\nu \propto \nu^\alpha$, where $S_\nu$ is the mean flux density at the
observing frequency $\nu$ and $\alpha$ is the spectral index \citep{Sieber+1973}.
However, the Vela pulsar presents a broken power-law spectral
form with a spectral index of $-0.55\pm0.03$ before and $-2.24\pm0.09$ after a
spectral break at $880\pm50$ MHz \citep{Jankowski+etal+2018}.
According to the recent measure of the mean flux density
($7\pm4$ mJy) at 5000 MHz \citep{Zhao+etal+2019}, the derived mean flux density from 
the power-law relationship at 6800 MHz is $3.58\pm2.11$ mJy, which is consistent with 
our measurement.

A pulse phase blind search algorithm was carried out to select
giant micropulses in the whole de-dispersed timestreams, which are required to
have signal-to-noise ratio (S/N) larger than 10.
The detection threshold corresponds to a limiting flux of 2.5 Jy.
This yielded a complete sample of 4187 giant micropulses were detected, most of which
are close to the detection threshold.

\section{Results}
\label{sec:results}

\begin{figure}
	\centering
	\includegraphics[width=8.0cm,height=6.0cm,angle=0]{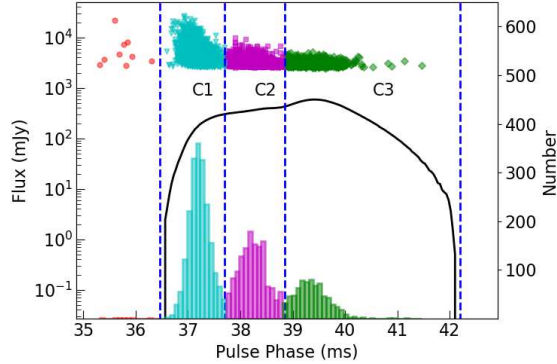}
	\caption{The folded pulse profile (black solid curve) and the giant micropulses
	as a function of pulse phase.
	The boundaries are plotted with blue dashed lines to distinguish the three
	main emission regions.
	It is noted that three clusters of giant micropulses (indicated with cyan,
	magenta and green points) are consistent with the three main emission
	components shown in the average pulse profile.
	Nine giant micropulses indicated with red points are confined to the leading
	edge of the profile.
	The histogram of giant micropulse distribution in pulse
	phase is plotted, and three clusters are indicated with corresponding colours.
	The fluxes of the giant micropulses are as measured in the 40.96 $\mu$s binned
	timestream used to find them.}
	\label{pic:profile}
\end{figure}

Figure~\ref{pic:profile} shows where the giant micropulses arrive relative to the
average pulse profile.
Three main emission regions correspond to the leading, central
and trailing components of the average pulse profile, and the phase boundaries
are shown with vertical dashed lines.
The giant micropulse emission are present at a wide phases, not just prior to
the main pulse window as reported by \citet{Johnston+etal+2001}.
Three clusters are presented with four giant micropulses falling in the trailing
edge of the pulse profile, which are consistent with three main emission regions.
It is worthy to note that nine giant micropulses appear at the pulse phase prior
to the nominal main emission window.
The maximum phase jitter between giant micropulses and the main pulse peak is
measured to be $\sim$4 ms, which is greater than that of 2.2 ms at 1413 MHz
\citep{Johnston+etal+2001}.
The brightest of these has a peak flux density in excess of 21.8 Jy, almost 40
times the peak flux density in the integrated pulse profile.
Very little overall effect on the integrated profile is resulted from these
giant micropulses.
This may give an indication of the nature of the pulse emission process.

\begin{figure*}
	\centering
	\includegraphics[width=8.0cm,height=10.0cm,angle=0]{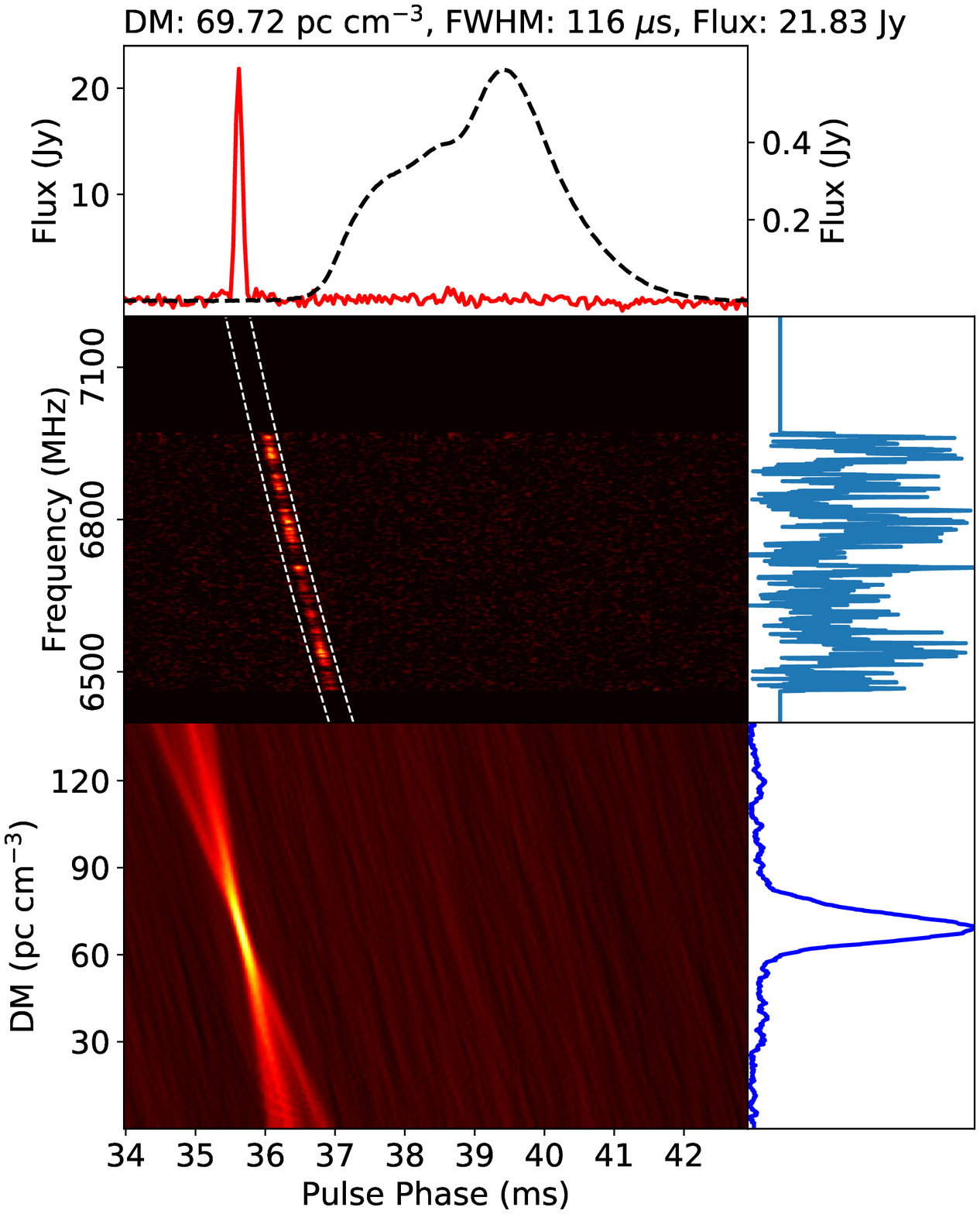}
	\includegraphics[width=8.0cm,height=10.0cm,angle=0]{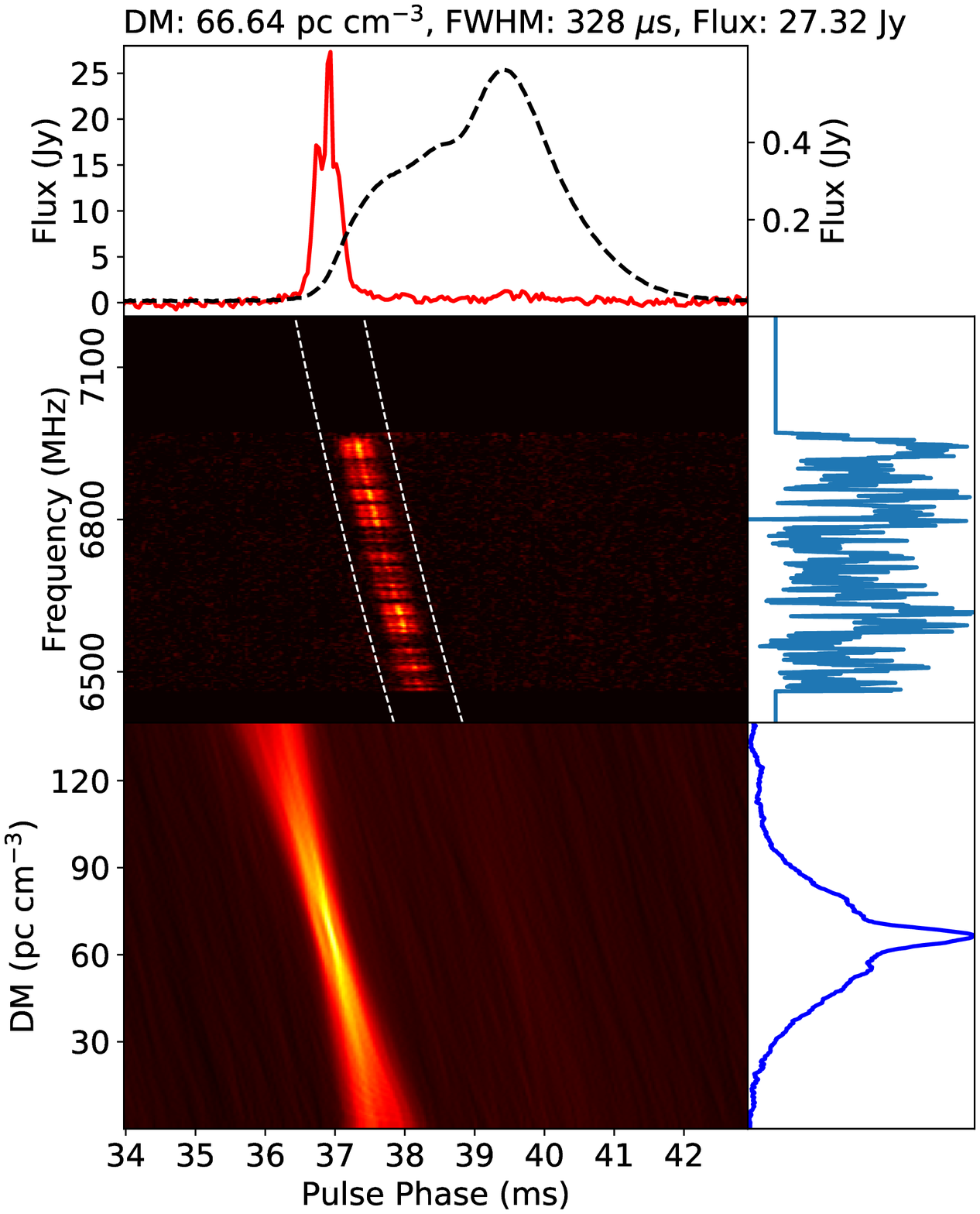}
	\includegraphics[width=8.0cm,height=10.0cm,angle=0]{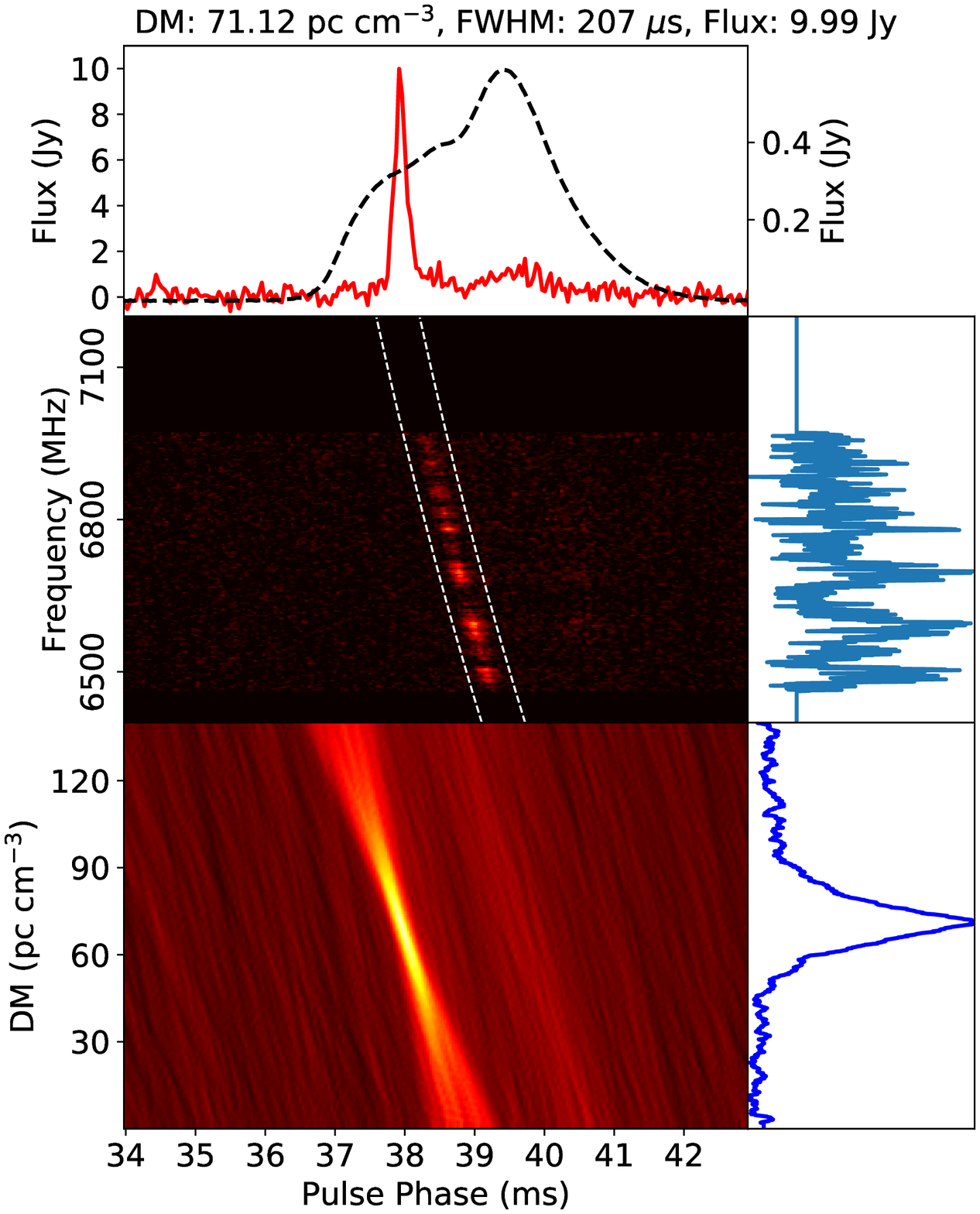}
	\includegraphics[width=8.0cm,height=10.0cm,angle=0]{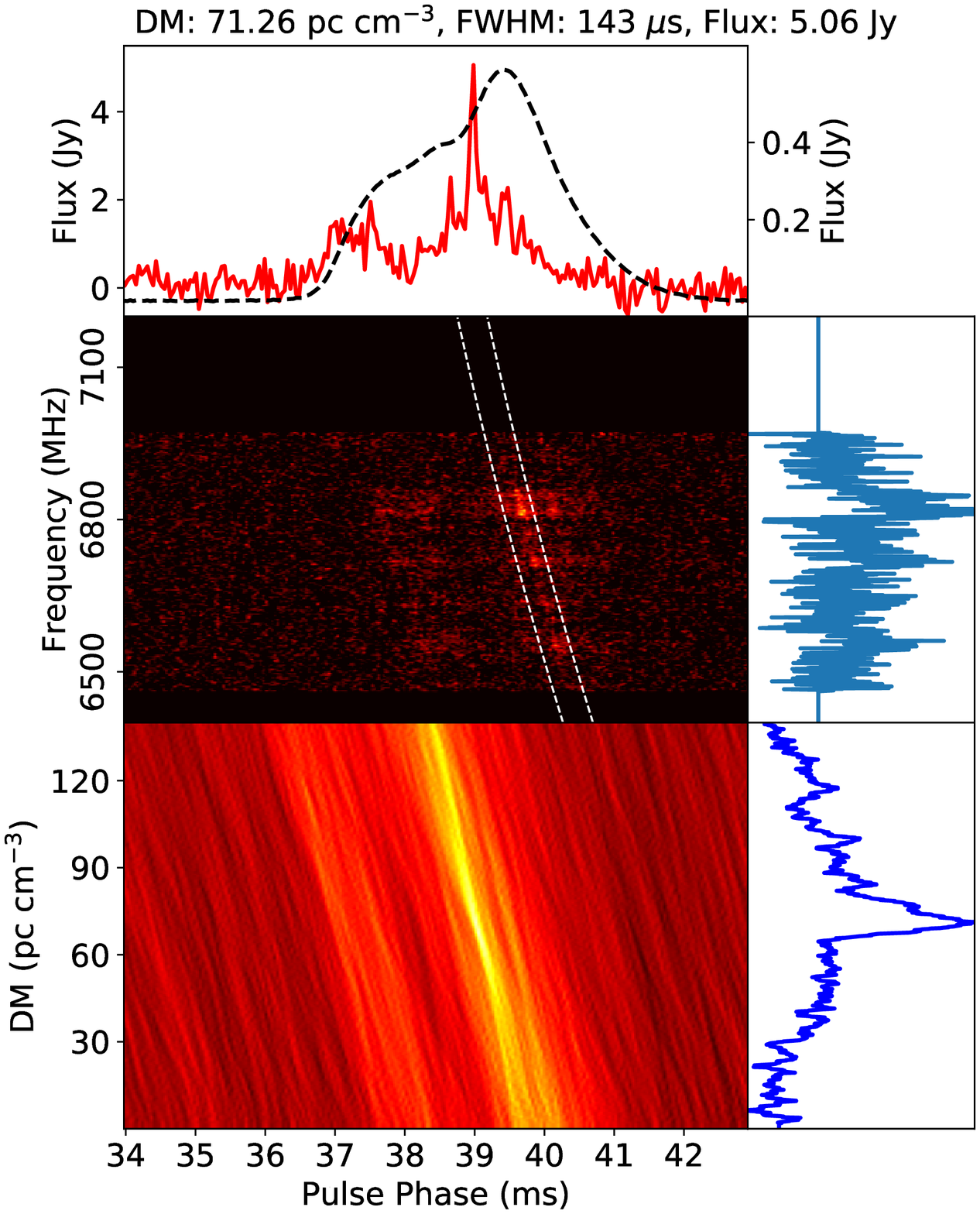}
	\caption{Examples of detected giant micropulses in different pulse phases.
	The upper panel is a time series graph showing the amplitude of the burst at
	DM=67.97 \pcm (red solid curve). 
	The averaged pulse profile is shown as black dashed line for comparison.
	Middle left panel: the time-frequency colour map shows the $\nu^{-2}$
	dispersive sweep of the burst. 
	The bandpass rolls off at the edge of the observing frequency.
	The dashed white lines illustrate the expected sweep for DM=67.97 \pcm.
	The de-dispersed spectra is projected to the middle right panel.
	Lower left panel: the DM-phase colour-coded diagram generated by
	de-dispersing the signal with DM varying from 0 to 140 \pcm in steps of
	0.14 \pcm.
	The lower right panel is a DM-S/N graph calculated from the maximum amplitude 
	over a whole pulse period.
	The measured DM, FWHM and peak flux density of the giant micropulse are
	listed on the title.}
	\label{pic:gp_profiles}
\end{figure*}

\begin{figure}
	\centering
	\includegraphics[width=8.0cm,height=6.0cm,angle=0]{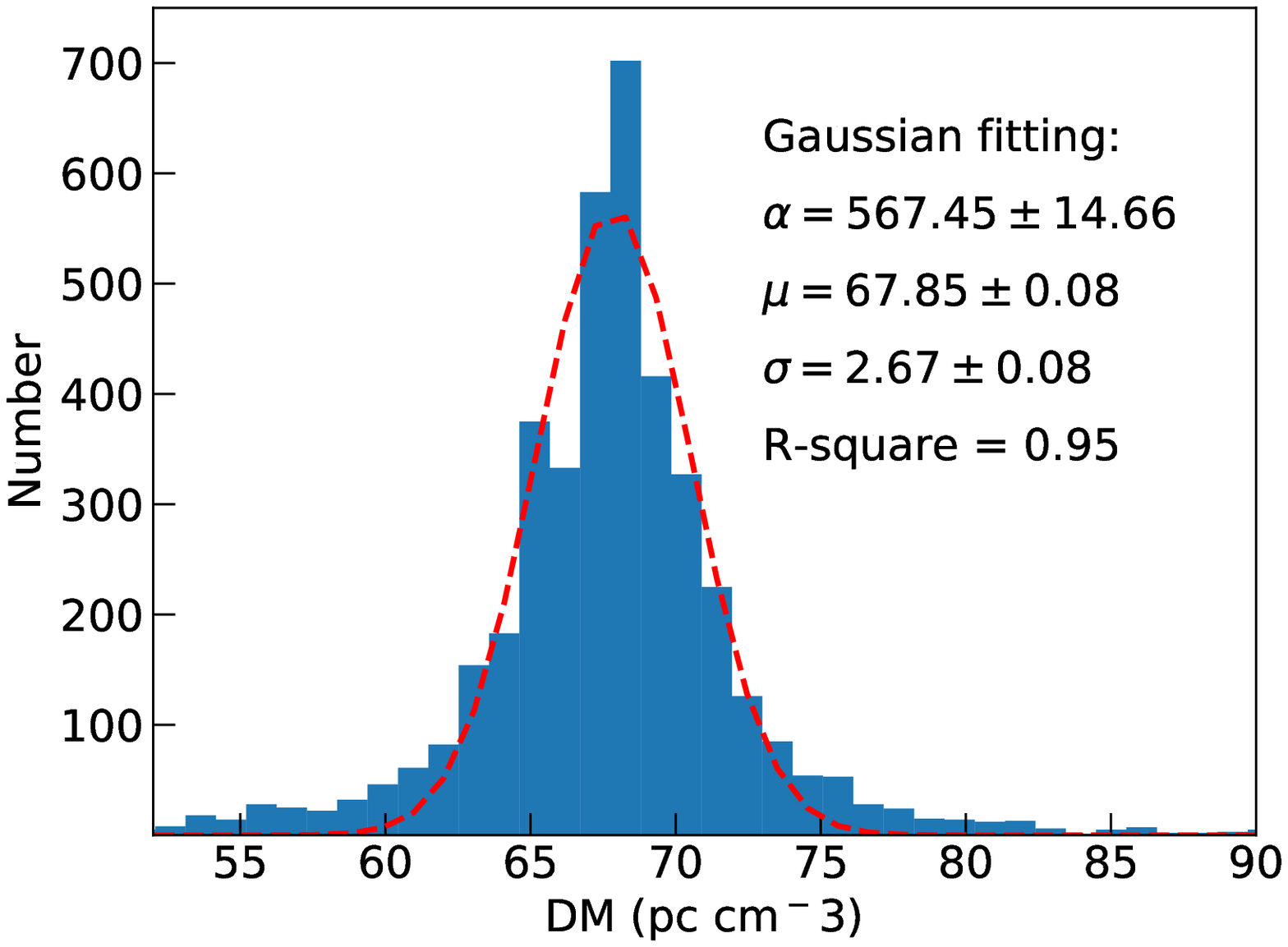}
	\caption{Histogram of the DM from the detected giant micropulses.
	The red dashed curve stands for the constrained optimal Gaussian
	distribution with the best fitting parameters shown in the text.}
	\label{pic:dm_dist}
\end{figure}

In order to identify whether the giant micropulses are originated from the Velar pulsar
or terrestrial interference, the selected signals with S/N above the detection
threshold are reprocessed with DM varying from 0 to 140 \pcm\ in steps of 0.14
\pcm.
To each DM, the time series after de-dispersion are shown in the lower left
panels of Figure~\ref{pic:gp_profiles}.
The burst dissolves as the DM increases or decreases from the nominal DM of
67.97 \pcm, which provides a significant criterion for identifying a real burst.
We are confident that the giant micropulses presented are not influenced by spurious
signals after visual inspection, since no sources of interference are seen with
dispersion like that of the Vela pulsar.
Then the DM of a giant micropulse is calculated from the maximum
amplitude over the whole pulse period (shown in the lower right panels).
The middle left panels present four example dynamic spectra of
the detected pulses.
The $\nu^{-2}$ dispersive sweep of the burst is clearly shown in the patchy
spectra, which further demonstrates the authenticity of the pulses.
Figure~\ref{pic:dm_dist} shows the histogram for the DMs derived
from all giant micropulses. 
The mean value of DM is 67.39$\pm6.26$ \pcm, which is in agreement with the value
published by \citet{Petroff+etal+2013}.

\begin{figure}
	\centering
	\includegraphics[width=8.0cm,height=6.0cm,angle=0]{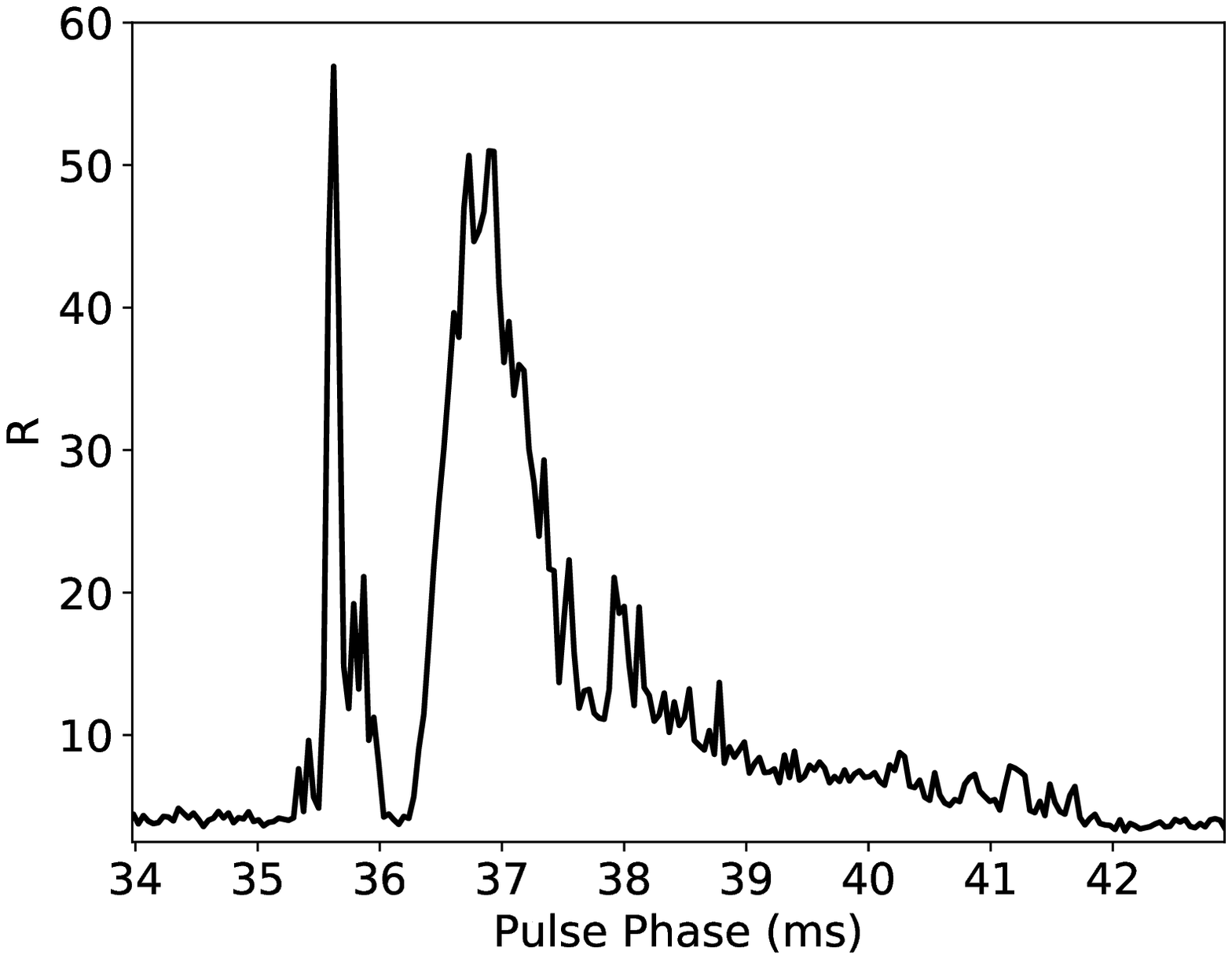}
	\caption{R-parameter as a function of pulse phase at 6800
	MHz.}
	\label{pic:r_param}
\end{figure}

As reported by \citet{Johnston+etal+2001}, a threshold of $R$ 
larger than 25 was taken to pick out the giant micropulses at 1413 MHz.
The $R$ parameter is defined as $R_i=(MAX_i-m_i)/\sigma_i$, where $MAX_i$ is the 
maximum intensity, $m_i$ is the mean intensity and $\sigma_i$ is the rms in the 
$i$th bin.
The phase-resolved $R$ at 6800 MHz is shown in Figure~\ref{pic:r_param}.
Two peaks are clearly presented in the leading and leading edge of the pulse,
which implies the existence of giant micropulses with extremely high amplitude
with respect to the mean intensity in these phase ranges.
And the value of $R$ decreases exponentially in the center and trailing of the
pulse, which is inconsistent with the Gaussian statistics at 1413 MHz.
As shown in Figure~\ref{pic:profile}, the giant micropulse emission in the
center component have higher occurrence rate and higher amplitude than that in
the trailing component.
Furthermore, none detection of giant micropulses with significant $R$-value 
occur in the bump region.

\begin{figure}
	\centering
	\includegraphics[width=9.0cm,height=8.0cm,angle=0]{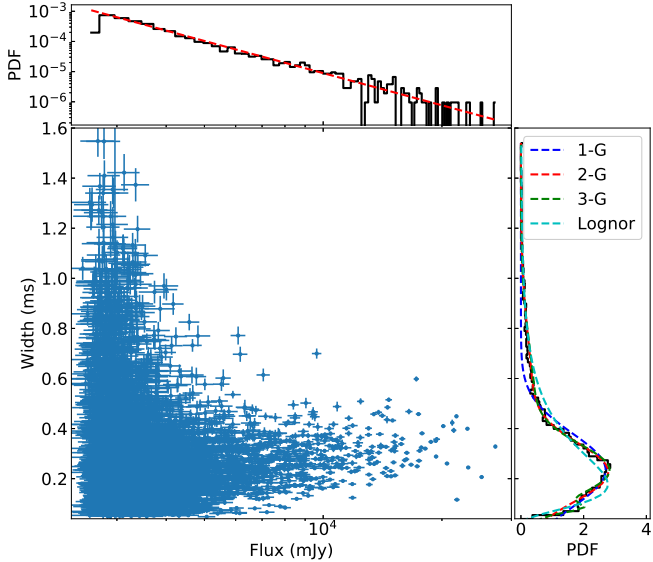}
	\caption{Lower left panel: scatter plot of width and peak flux of 4187 giant
	micropulses.
	Upper left panel: PDF of peak flux densities along with the best fit power law
	probability distribution corresponding to $\alpha = -3.54\pm0.04$ in
	log-log space.
	The most energetic pulse in our sample has a peak flux density $\sim$28Jy.
	Lower right panel: PDF of giant micropulse widths.
	The majority of giant micropulses tend to cluster in width of 50 to 500 $\mu$s.
	The best fits with one, two and three Gaussians and a log-normal 
	distributions are indicated with blue, red, green and cyan dashed lines, respectively.}
	\label{pic:gp_distribution}
\end{figure}

\begin{figure}
	\centering
	\includegraphics[width=8.0cm,height=6.0cm,angle=0]{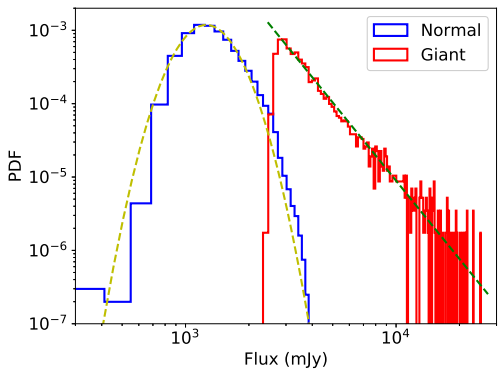}
	\caption{PDF of the peak flux densities for the normal
	pulses (blue) and detected giant micropulses (red) from the Vela pulsar.
	The yellow dashed line shows the expected logarithmic normal distribution
	which fits the PDF for the normal pulses well.
	While the PDF for the giant micropulses can be well described by a power-law
	distribution.}
	\label{pic:spectra}
\end{figure}

\begin{figure}
	\centering
	\includegraphics[width=8.0cm,height=6.0cm,angle=0]{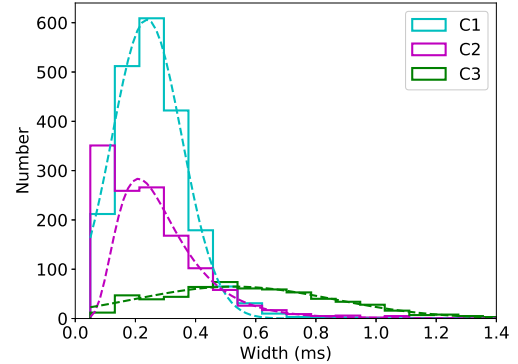}
	\caption{Histograms of pulse widths for the three phase clusters of detected
	giant micropulses.
	The expected normal distributions for C1 and C3 are shown with dashed curves
	in corresponding colours.
	While, the histogram for C2 has a rough lognormal distribution.}
	\label{pic:width}
\end{figure}

The main panel of Figure~\ref{pic:gp_distribution} shows a scatter plot of peak
flux density and the full width of half maximum (FWHM) for detected giant micropulses.
The pulse broadening caused by inner-channel dispersion (1.4
$\mu$s) and scattering (1.0 ns) is negelected due to high observing frequency.
To obtain the FWHM, a Gaussian function is adopted to fit each giant 
micropulse.
The position of the center of the peak is kept fixed with the pulse phase of the
maximum value.
Then FWHM is given by $\rm{2\sqrt{2ln(2)}} \sigma$, where $\sigma$ is the
determined parameter using the least-square method.
The giant micropulses have timescales ($<$1.55 ms) much smaller than that of the average
pulse profile (2.62 ms).
As can be seen, the majority of giant micropulses tend to cluster in width of 50
to 500 $\mu$s.
In this interval, the width of a giant micropulse seems to be independent of its
peak flux density, which is consistent with the result at 2.3 GHz \citep{Kramer+etal+2002}.
The pulse-to-pulse energy distribution is served as one of the
important differentiator between various emission processes
\citep{Burke-Spolaor+etal+2012}.
Probability density function (PDF) for the peak flux densities is shown in
the upper panel, which is well described by a power law.
The slope, obtained via least-square fitting, is give by $\alpha =
-3.54\pm0.04$, which is the steepest distribution to our best knowledge.
While a logarithmic normal distribution is presented for the PDF formed from the
peak flux densities of normal pulses, as shown in Figure~\ref{pic:spectra}.
The distinctive distributions between giant micropulses and normal pulses may
indicate their different emission mechanism.
The pulse energies of the bursts from the Vela pulsar do not exceed 10 times the
corresponding mean quantity.
Nevertheless, the peak flux densities of the bursts are very large in absolute
terms.
For example, the brightest pulse detected corresponds to a peak flux density of
approximately 28 Jy, which is 46 times the peak flux density of averaged
profile.
The pulse width PDF is given in the right panel, which clearly shows a normal
distribution centered at around $\sim$250 $\mu$s.
In order to test if the pulse width PDF is multimodal for FWHM$>50\ \mu$s, we 
model it as a sum of Gaussian distributions and a lognormal distribution.
Models composed of one and up to three Gaussians and a lognormal distribution are
tested against the data.
The best fits obtained are presented in the right panel of
Figure~\ref{pic:gp_distribution}.
Table~\ref{tab:aic} gives the parameters of the models using the
maximum likelihood technique.
The lognormal distribution gives the best description of the
pulse width PDF using the Akaike information criterion
\citep[AIC,][]{Akaike+1974}.
We note, however, that distinctive distributions are presented for the phase
resolved pulse width distributions as shown in Figure~\ref{pic:width}.
The histograms of pulse widths for the first and third clusters both show normal 
distributions, where the third cluster has a greater typical width than the
first cluster.
While, a lognormal distribution is presented for the pulse width histogram for the
second giant micropulse cluster.
The best-fit values of the amplitude $\alpha$, the mean $\mu$ and
the standard deviation $\sigma$ of normal distributions for C1 and C3 and a
lognormal distribution for C2 are listed in Table~\ref{tab:width}.

\begin{table}
	\centering
	\caption{Results of the fits and the AIC applied to the 
	pulse width PDF modeled as a sum of Gaussian components and a lognormal
	distribution.}
	\label{tab:aic}
	\begin{tabular}{ccccc}
		\hline
		\hline
		Parameters & 1-G & 2-G & 3-G & Lognormal \\
		\hline
		$\alpha_1$ & 1     & 0.73  & 0.66  & 1 \\
		$\mu_1$    & 0.23  & 0.22  & 0.23  & -1.30 \\
		$\sigma_1$ & 0.15  & 0.12  & 0.10  & 0.66 \\
		$\alpha_2$ & ...   & 0.27  & 0.04  & ... \\
		$\mu_2$    & ...   & 0.48  & 0.08  & ... \\
		$\sigma_2$ & ...   & 0.28  & 0.02  & ... \\
		$\alpha_3$ & ...   & ...   & 0.30  & ... \\
		$\mu_3$    & ...   & ...   & 0.46  & ... \\
		$\sigma_3$ & ...   & ...   & 0.27  & ... \\
		R-square   & 0.93  & 0.96  & 0.98  & 0.95 \\
		$AIC$      & 888.49& 198.64& 208.80&168.88\\
		$\Delta_n$ & 719.61& 29.76 & 39.92 & 0    \\
		$\omega_n$ & $10^{-157}$ & $10^{-7}$ & $10^{-9}$ & $\sim$1 \\
		\hline
	\end{tabular}
	\begin{flushleft}
		\textbf{Notes.} The label $n$-G in each column denotes the model of
		a sum of $n$ Gaussians.
		$\alpha$, $\mu$ and $\sigma$ are the weight, mean and the standard
		deviation of the component of the sum of Gaussians.
		$AIC$, $\Delta_n$ and $\omega_n$ represent the AIC value, the relative
		$AIC$ respect to the model with the minimum $AIC$ value, and the Akaike
		weights of the model, respectively.
		R-square stands for the value of goodness of fit.
	\end{flushleft}
\end{table}

\begin{table}
	\centering
	\caption{The best-fit parameters for the pulse widths of three
	phase clusters. 
	The C1 and C2 are fitted with normal distributions, and C3 is fitted with a
	lognormal distribution.}
	\label{tab:width}
	\begin{tabular}{cccc}
		\hline
		\hline
		Parameters & $\alpha$ & $\mu$ & $\sigma$ \\
		\hline
		C1 & 177$\pm$3 & 0.238$\pm$0.002 & 0.116$\pm$0.002 \\
		C2 & 79.5$\pm$0.9 & -1.336$\pm$0.002 & 0.476$\pm$0.007 \\
		C3 & 54$\pm$3 & 0.53$\pm$0.02 & 0.33$\pm$0.02 \\
		\hline
	\end{tabular}
\end{table}

\begin{figure}
	\centering
	\includegraphics[width=8.0cm,height=6.0cm,angle=0]{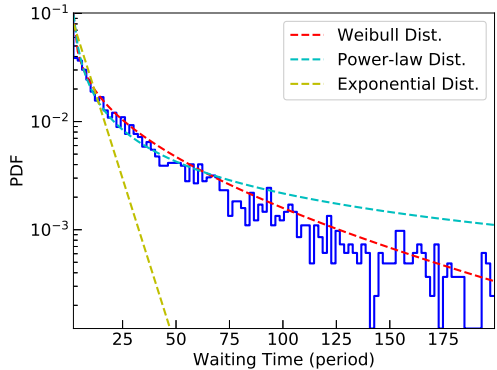}
	\caption{Waiting time distributions of detected giant micropulses (blue
	solid line).
	The best-fit Weibull (red), power-law (cyan) and exponential (yellow) 
	distributions are presented with dashed lines, respectively.}
	\label{pic:WTD}
\end{figure}

In order to investigate whether a giant micropulse is related to or independent of the
previous one, the intervals between successive giant micropulses (waiting time, 
$\Delta t$) are calculated.
The statistics of waiting time is intensively studied for solar flares, which
can provide critical information about how an individual event occurs
\citep{Wheatland+2000}.
The waiting time PDF shown in Figure~\ref{pic:WTD} presents a domination of short
waiting times ($<$25 periods), which indicates the production of giant micropulses 
occur in clusters, in other words, with a certain amount of memory.
It leads us to interpret the distribution of giant micropulse waiting time with
the Weibull distribution:
\begin{equation}
	P(\Delta t) = \frac{k}{\beta} (\frac{\Delta t-\theta}{\beta})^{k-1}
	e^{-(\frac{\Delta t - \theta}{\beta})^k},
\end{equation}
where $\beta$ is the reciprocal of the mean occurrence rate.
Using the maximum-likelihood estimation, the best fitting coefficients are 
$k = 0.66\pm0.01$, $\beta = 28.71\pm1.09$, $\theta = 1.69\pm0.05$ and R-square = 0.99.
A $k < 1$ implies that the probability of a giant micropulse
occurring decreases with time, namely, the giant micropulse occurrence is
clustered.
A $\theta > 0$ describes that the occurrence probability is zeros for consecutive giant
micropulses.
The estimated occurrence rate is higher than that calculated from our
observation, which appears to be resulted from the clustering effect.
A simple Poisson process, where the probability of a giant micropulse occurring 
is time invariant, produces an exponential distribution, which is fitted poorly
as shown in the yellow dashed line.
The best-fitted power-law distribution with an index of $-0.97\pm$0.02 does not
fit the waiting time distributions well, as shown in Figure~\ref{pic:WTD}.

\section{Discussion}
\label{sec:discussion}

The origin of giant pulses has been remaining a mystery since the discovery of
giant pulses from Crab pulsar \citep{Staelin+Reifenstein+1968}.
The generation of giant pulse activity was pointed to be an intrinsic phenomenon
within the pulsar \citep{Hankins+1971}.
The Giant pulses are supposed to be the product of induced Compton scattering of 
the radio radiation off the plasma in the pulsar magnetosphere \citep{Petrova+2006}.
The extremely high intensity is as well caused by an enhanced number of charges
partaking in the nonthermal, coherent radiation processes
\citep{Hankins+etal+2003}.
Alternatively, the origination of giant pulses is proposed from the coherent 
instability of plasma near the magnetic equator of light cylinder \citep{Wang+etal+2019}.
\citet{Singal+Vats+2012} suggested that the giant pulse emission and nulling may 
be opposite manifestations of the same physical process.
The giant pulses are suggested to occur in pulsars with
extremely high magnetic fields at the light cylinder of $B_{LC} > 10^5$ G
\citep{Cognard+etal+1996}.
Therefore, the giant pulses are proposed to be originated near the light
cylinder \citep{Istomin+2004}.
However, the giant pulses are also detected in the pulsars with ordinary
magnetic fields at the light cylinder of $B_{LC} < 100$ G, such as PSRs 
B0031$-$07 \citep{Kuzmin+Ershov+2004}, B1112+50 \citep{Ershov+Kuzmin+2003}, 
J1752+2359 \citep{Ershov+Kuzmin+2005}, B0950+08 \citep{Smirnova+2012}. B0656+14
\citep{Kuzmin+Ershov+2006}, B1237+25 \citep{Kazantsev+Potapov+2017} and B0301+19
\citep{Kazantsev+etal+2019}, and it does not seem to support the high $B_{LC}$ hypothesis.
The Vela giant micropulse emission physics maybe independent on the high 
magnetic field at the light cylinder.
Although the Vela's $B_{LC}$ is about 20 times smaller than that of PSR B1937+21
and the Crab pulsar, it is still in the top 5\% of pulsars with $B_{LC}$ estimate.
The giant pulses from PSR J1824$-$2452A occur in narrow phase windows that
correlate in phase with X-ray emission, and the two emission phenomena likely
originate from the similar magnetospheric regions but not the same physical
mechanism \citep{Knight+etal+2006}.
In order to reveal the nature of the giant micropulses, simultaneous radio and
X-ray observations on the Vela pulsar will be required.

Considering a scenario of narrow band emission, the high-frequency emission is
assumed to be generated at low altitude and vice verse, called
radio-to-frequency mapping (RFM) \citep{Cordes+1978}.
This empirical relationship is well demonstrated from the Vela pulsar, the 
integrated pulse profile becomes narrower and narrower along with the 
increasing frequency \citep{Liu+etal+2019}.
Plausible interpretations are proposed for the two types of giant micropulse
emission.
The giant micropulses may originate from two different emission regions in the
pulsar magnetosphere.
A certain amount of memory shown in the occurrance of giant micropulses may
indicate that the normal giant micropulses are likely arised from the homologous
region with the normal pulse emission, but with a different plasma state, since 
they are coincident with averaged profile in pulse phase.
For instance, the fluctuations in the number of charges partaking in the coherent
radiation process that gives rise the intense variation in the net radio emission 
of the pulse intensity,
These kind of giant micropulses are emitted with high occurrance rate, because
the frequent turbulence of plasma in the inner acceleration region could result
in the enhancement of subbeam emission.
While the leading giant micropulses are likely originated from a higher altitude
within the same magnetic flux tube than the normal pulses.
In this model, the giant micropulses are supposed to be accompanied by normal pulse
emission, which are shown in some cases.
However, according to the RFM, the higher frequency the narrower pulse width,
which is inconsistent with the fact that our observed jitter in the arrival times 
at 6800 MHz is greater than that at lower frequencies.
Therefore, the giant micropulses at different frequencies may be emitted from
different magnetic field lines.
The giant micropulse emission region at higher frequency is closer to the last
open dipolar field lines than that at low frequency.

The Vela pulsar is known to be active in glitching, at least 7
glitches have been reported since 2003 
\footnote{http://www.atnf.csiro.au/research/pulsar/psrcat/}.
An intensive single-pulse observing campaign of the Vela
pulsar at 1376 MHz showed that the pulse profile varied temporally, and was 
affected with a micro-glitch \citep{Palfreyman+etal+2016}.
Furthermore, recent analysis of the 2016 glitch in the Vela pulsar,
the accompanying alteration of the magnetospheric was observed
\citep{Palfreyman+etal+2018}.
Therefore, the detailed phase distribution of giant micropulses
possibly provide additional clues on how the magnetosphere changes.
As suggested by \citet{Palfreyman+etal+2016}, the widening of emission 
cone could be caused by a glitch since the emission beam approaches the line of
sight.
Meanwhile, the giant micropulse distribution broadens under the assumption that
both regular and giant micropulses originate from the same emission region.
Futhermore, The glitch could lead to the increase of the plasma
density in open field lines. 
The coherent instability enhances due to plasma oscillation.
Then the giant micropulses within the emission window emerge, which broadens the
distribution.
The variation of pulse phase distribution of giant micropulses
after the 2016 glitch is the topic of continuing observations.
The radio spectrum of giant pulses shows a power-law distribution,
its spectral index is compared to the average pulse value for a pulsar
\citep{Popov+etal+2006}.
The giant pulses do not occur simultaneously in both frequency
ranges \citep{Popov+Stappers+2003}.
Our detected giant micropulse rate at 6800 MHz is 1/36, which is 
higher than the previous bright pulse rate at 1376 MHz.
The rate of bright pulse activity was reported to increase after 
some micro-glitches \citep{Palfreyman+etal+2016}.
Therefore, this increase in statistics is possible to be affected by the glitches.
The separation of giant pulse emission regions at lower
frequency is larger than that at higher frequency for PSR B0031$-$07 
\citep{Kuzmin+Ershov+2004}, which is contrary to the Vela pulsar.
Therefore, the temporal evolution is preferrable to cause the
observed difference in the pulse phase distribution of giant micropulses.

Further long-term simultaneous multi-frequency single pulse observations with full
Stokes parameters would be very worthwhile in discerning the pulse emission
mechanism and glitching process.

\section*{Acknowledgements}
\addcontentsline{toc}{section}{Acknowledgements}
We thank the referee for comments which improved the paper.
Much of this work was made possible by grant support from the West Light Foundation 
of Chinese Academy of Sciences (WLFC 2016-QNXZ-B-24), the Chinese National 
Science Foundation Grant (U1838109, U1731238, U1831102, U1631106, 11873080), 
the National Basic Research Program of China (973 Program 2015CB857100).
JPY is supported by a prospective project of the Astronomical Research Center of
the Chinese Academy of Sciences.
NW is supported by the National Program on Key Research and Development Project
(grant No. 2016YFA0400804).
HGW is supported by 2018 project of Xinjiang uygur autonomous
region of China for flexibly fetching in upscale talents.
JLC is supported by the Scientific and Technologial Innovation Programs of Higher 
Education Institutions in Shanxi (Grant No. 2019L0863).
We thank members of the Pulsar Group at XAO for helpful discussions.
We thank the staff of the Yunnan Astronomical Observatory who have made these 
observations possible.

\bibliographystyle{aasjournal}
\bibliography{gpbibtex}

\end{document}